\documentclass{mem}
\usepackage{natbib}
\usepackage{txfonts}
\usepackage{balance}
\usepackage{graphicx}
\usepackage[a4paper]{hyperref}
\idline{79}{1}
\begin{document}
\title{Observational constraints on AGB mass loss and its effect on AGB
evolution}
\subtitle{}
\author{Jacco Th. \,van Loon\inst{1}}
\institute{Lennard-Jones Laboratories, Keele University, Staffordshire ST5
5BG, United Kingdom\\
\email{jacco@astro.keele.ac.uk}}
\authorrunning{van Loon}
\titlerunning{AGB mass loss}
\abstract{
This review discusses some of the observational constraints on what we know
about the mass loss experienced by stars in the Asymptotic Giant Branch (AGB)
phase of evolution. Mass loss affects the maximum mass attained by the core of
an AGB star and hence its fate as a white dwarf or potentially a supernova.
The way mass loss depends on stellar initial parameters and time affects the
yield from AGB stars, in terms of elemental abundances and types of dust. The
r\^ole of pulsation, dust formation, chromospheres and other mechanisms which
may contribute to mass loss are assessed against observational evidence, and
suggestions are made for observations which could force significant new
progress in this field in the first decades of the $21^{\rm st}$ century. A
better understanding of AGB mass loss may be gained from a combination with
studies of first ascent red giant branch (RGB) stars and red supergiants,
through population studies and in different environments.
}
\maketitle{}

\section{Evidence for mass loss on the AGB}

\citet{Shklovsky56} proposed an origin of the expanding Planetary Nebulae
(PNe) at the Asymptotic Giant Branch (AGB), the last phase of evolution of
stars with initial masses between $M_{\rm init}\sim0.8$ and 8 M$_\odot$.

\citet{Deutsch56} presented the detection of optical absorption lines in front
of the warm companion to the M5\,III giant in the $\alpha$\,Herculi binary. He
interpreted this as a wind emanating from the cool giant, which loses mass at
an estimated rate $\dot{M}>3\times10^{-8}$ M$_\odot$ yr$^{-1}$.

\citet{Gehrz71} detected infrared (IR) emission from warm --- hence
circumstellar --- dust around cool giant stars, including $\alpha$\,Her for
which they estimated $\dot{M}\sim9\times10^{-8}$ M$_\odot$ yr$^{-1}$,
consistent with Deutsch' analysis. For strongly pulsating (Mira) red giants
they derived much higher mass-loss rates, $\dot{M}\sim2\times10^{-6}$
M$_\odot$ yr$^{-1}$. They suggested that the winds of Miras may in fact be
driven through radiation pressure on these dust grains.

Hydroxyl (OH) maser emission is observed at radio wavelengths from the most
dust-enshrouded M-type stars (OH/IR stars). The wind velocity can be measured
from the line profiles, confirming the dust-driven wind scenario
\citep{Elitzur76,Richards98}. Rotational lines of the abundant carbon monoxide
molecule (CO) are present in emission at (sub)mm wavelengths in all dusty
winds including those of carbon stars \citep{Knapp85}. These are thermally
excited transitions and thus more reliable for mass-loss rate measurements.

\section{Global constraints on mass loss}

\subsection{Initial-final mass (IFM) relation}

AGB stars are believed to end their lives as a white dwarf, their
electron-degenerate carbon-oxygen core. These can not be more massive than
$\sim1.4$ M$_\odot$, or they would implode \citep{Chandrasekhar31}. If this
happens whilst still inside the AGB star, a supernova type 1.5 would result
\citep{Iben83}. This has never been seen to occur. Hence AGB stars with masses
$M_{\rm init}>1.4$ M$_\odot$ {\it must} shed their mantle to truncate any
further growth of their core. For the most massive AGB stars this means
several solar masses must be lost, and quickly enough.

The white dwarf mass distribution peaks at $\sim0.6$ M$_\odot$, both in the
Small Magellanic Cloud, SMC \citep{Villaver04}, in the Large Magellanic Cloud,
LMC \citep{Villaver07}, and in the Milky Way. The initial-final mass (IMF)
relation seems to vary little between galactic open clusters of different
metallicity \citep{Williams07}. This suggests that low-mass AGB stars must
also lose mass, and that the total mass lost on the AGB depends little, if
any, on the metal content. Differences within 0.1 $M_\odot$ can easily be
explained by differences in star formation histories or observational bias.

Does this also mean that the mass-loss {\it rate} depends little on
metallicity? No. AGB stars reach mass-loss rates two-three orders of magnitude
higher than the core growth rate \citep{Vanloon99b}. This causes the core to
grow very little during the phase of heaviest mass loss (this is not the case
for red supergiants, RSGs, which seem unable to escape a supernova ending). If
the mass-loss rates were lower by an order of magnitude, AGB stars would live
longer, but not enough for the core to grow a lot. Only when the mass-loss
rate drops to within a few times the core growth rate will the core grow
significantly. This would require the mass-loss rate not to exceed
$\sim10^{-7}-10^{-6}$ M$_\odot$ yr$^{-1}$ for any significant amount of time
(depending on core mass), unlike what is observed in the Magellanic Clouds
\citep{Vanloon99b} and solar neighbourhood \citep{Jura89}.

\subsection{Population studies and yields}

For all but the most extreme initial mass functions, any stellar population
will eventually produce AGB stars. Population studies of the luminosity
functions of oxygen-rich and carbon AGB stars \citep{Groenewegen93} could, in
principle, constrain AGB mass-loss, but these studies usually adopt a certain
mass-loss formalism, concentrating instead on calibrating aspects of internal
processes such as the dredge-up efficiency. But for example an apparent lack
of PN precursors and luminous carbon stars \citep{Wood83,Reid90} was
alleviated after IR surveys revealed a large population of optically invisible
stars in a phase of intense mass loss and dust production
\citep{Wood92,Vanloon97,Vanloon06b}.

The AGB stars will lose mass enriched in products of nucleosynthesis, and
dust. These products are encountered in the interstellar medium (ISM), next
generations of stars or even in our Solar System, and can thus shed light on
AGB mass-loss.

\citet{Ferrarotti06} showed that different mass-loss formalisms result in
different timings during AGB evolution of the Mira phase and subsequent OH/IR
phase, and that dust production depends on initial metallicity. They predicted
that at solar metallicity, Z$_\odot$, a 2 M$_\odot$ star produces little
carbonaceous dust, as it becomes a carbon star only very late, whilst stars of
low initial metallicity, $Z_{\rm init}=\frac{1}{20}$ Z$_\odot$, produce solely
carbonaceous dust across the entire AGB mass range.

\citet{Zinner06} analysed meteoritic evidence for the origin of dust. They
studied the silicon carbide grains due to AGB carbon stars, most of which
appear to come from metal-rich stars. They require a rather low mass-loss
efficiency, Reimers' Law \citep{Reimers75} with efficiency $\eta=0.1$, to
explain the observed isotopic ratios. Stronger mass loss results in fewer
thermal pulses on the AGB, affecting the chemical yields. \citet{Karakas06}
found that adopting Reimers' Law instead of the \citet{Vassiliadis93}
prescription causes $\sim75$\% drop in yields for elements such as magnesium,
aluminium, and silicon.

\section{What is the rate of AGB mass-loss?}

\subsection{Measured mass-loss rates}

The most common methods to derive mass-loss rates from AGB stars are based on
the IR emission from the circumstellar dust, or CO emission from the molecular
envelope; they were reviewed recently in \citet{Vanloon07b} and
\citet{Schoeier07}, respectively.

A compilation of mass-loss rates from Galactic M-type, carbon and the
intermediate S-type (carbon:oxygen ratio nearly unity) AGB stars was published
recently by \citet{Guandalini06} and \citet{Busso07}. The carbon stars were
found to be the more obscured stars in general, but this is in part due to the
higher opacity of carbonaceous dust. They reach $\dot{M}\sim10^{-5}$ M$_\odot$
yr$^{-1}$ \citep{Jura90}, similar to OH/IR stars \citep{Olnon84} but much
higher than the less evolved optically bright Miras \citep{Jura92}. The S-type
stars were initially found to have comparatively low mass-loss rates,
$\dot{M}\sim10^{-7}$ M$_\odot$ yr$^{-1}$ and a dust:gas mass ratio, $\psi$, a
few times lower than the typical $\psi\sim0.005$ in M-type and carbon stars
\citep{Jura88}. But \citet{Ramstedt06} find a similar median
$\dot{M}\sim2\times10^{-7}$ M$_\odot$ yr$^{-1}$ for S-type, M-type and carbon
AGB stars on the basis of CO emission. It is therefore not clear whether the
mass-loss rate depends critically on the carbon:oxygen ratio.

Complete samples of AGB stars may be obtained in the Magellanic Clouds, and
their mass-loss rates appear to reach similar values as in the Milky Way after
scaling the wind speed and dust:gas ratio with metallicity and luminosity
\citep{Vanloon00,Vanloon06}. This was confirmed recently in other nearby dwarf
galaxies \citep{Jackson07a,Jackson07b,Lagadec07b}. \citet{Groenewegen07},
using {\it Spitzer Space Telescope} spectra, also obtained similar mass-loss
rates for SMC and LMC carbon stars, but using Galactic values for the wind
speed and dust:gas ratio; they also find an identical correlation with period
of pulsation as in the Milky Way. It must be noted that the sample is heavily
restricted to carbon stars only, in a narrow mass range around $\sim1.5$
M$_\odot$.

\citet{Blommaert06} modelled the spectral energy distributions of M-type AGB
stars in the Galactic Bulge, deriving low mass-loss rates,
$\dot{M}\sim10^{-8}-10^{-7}$ M$_\odot$ yr$^{-1}$. These are low-mass stars,
$M_{\rm init}\sim1$ M$_\odot$, and it must be realised that in smaller stellar
systems it is more difficult to capture the brief phase of intense mass-loss.
This is in particularly true for star clusters \citep{Jura87,Vanloon05b}.

The slow winds and ability to detect cold dust and gas allow variations in the
mass-loss rate to be traced over as much as $10^4$ yr, about a thermal-pulse
interval. \citet{Jura86} and \citet{Groenewegen07} found $>10$\% of carbon
stars to have varied noticeably in mass-loss rate over the past $10^2-10^3$
yr. \citet{Decin07} presented a detailed account of mass-loss variations in an
OH/IR star, WX\,Piscis.

\subsection{One formula fits all?}

An early empirical formula for the mass-loss rates of red giants, Reimers'
Law \citep{Reimers75} does not reproduce the very high rates found in OH/IR
stars and their carbon star equivalents. In a heroic attempt to describe the
mass loss across the entire Hertzsprung-Russell diagram with one
$\dot{M}(L,T_{\rm eff})$ formula, \citet{Dejager88} failed on the AGB.

\citet{Judge91} fitted $\dot{M}\propto g^{-1.5}$, where $g$ is the gravity, to
within an order of magnitude of data ranging from first-ascent red giant
branch (RGB) stars with $\dot{M}\sim10^{-9}$ M$_\odot$ yr$^{-1}$, to AGB stars
with $\dot{M}>10^{-5}$ M$_\odot$ yr$^{-1}$.

Vassiliadis \& Wood (1993) fit $\log\dot{M}\propto P$ to CO mass-loss rates
and pulsation periods, $P$, with no difference between M-type and carbon AGB
stars. Their sample is dominated by stars in a narrow range of mass, $\sim1-2$
M$_\odot$. The relation is shifted to longer periods at higher mass; this
introduces orders of magnitude difference in mass-loss rate at a given
pulsation period for stars that differ in mass by only a factor two or three.
At approximately $P>600$ days, the mass-loss rates of OH/IR stars derived from
the 60 $\mu$m flux density are seen to saturate around a value
$\dot{M}\sim10^{-4}$ M$_\odot$ yr$^{-1}$, just a little above the
single-scattering limit \citep{Jura84}.

A formula of the form $\dot{M}(L,T_{\rm eff})$ was derived by
\citet{Vanloon05} for oxygen-rich massive AGB stars and red supergiants (RSGs)
in the LMC. It was found to also describe Galactic stars, except some of the
warmer or weaker pulsating ones. The formula is similar to results from
hydrodynamical computations \citep{Wachter02} for carbon stars, which do
exhibit a similar behaviour in the LMC data: the mass-loss rate scales roughly
in proportion to luminosity, which provides the radiation pressure, and
strongly with lower temperature, which allows dust formation.

\section{How is AGB mass-loss driven?}

\subsection{The r\^ole of dust}

In the wind of a luminous, cool star, dust is always observed. That it also
drives the wind is confirmed with interferometric maser observations of nearby
OH/IR stars \citep{Richards98} and the wind speeds in OH/IR stars
\citep{Marshall04} and optical depths in both M-type AGB and carbon AGB stars
\citep{Vanloon00} in the Magellanic Clouds and Milky Way. \citet{Habing94}
predicted a slightly steeper luminosity dependence of the wind speed (exponent
0.3 instead of $\frac{1}{4}$), which would indeed fit better the data in
\citet{Marshall04}.

The Eddington luminosity required to drive a wind via dust is reached near the
tip of the RGB for carbonaceous dust, but only higher up the AGB for
silicates because they are more transparent \citep{Ferrarotti06}. The exact
threshold in terms of the coupling between the dust and gas fluids is
uncertain: \citet{Netzer93} estimated $\dot{M}>10^{-7}$ M$_\odot$ yr$^{-1}$
but \citet{Gail87b} $>10^{-6}$ M$_\odot$ yr$^{-1}$. To produce a dust-driven
wind from a pulsating M-type star, Woitke (2006) required iron to provide
opacity; this could help explain the metallicity dependence of AGB mass-loss
\citep{Vanloon06}.

Alternatively, \citet{Hoefner07} suggested that an M-type star may form some
carbonaceous dust, which could supply the opacity. This scenario could help
explain that the smooth transition from slow winds and low mass-loss rates to
faster denser winds measured in CO is undistinguishable between M-type, S-type
and carbon AGB stars \citep{Knapp98}.

There is a great deal uncertainty about the details of the dust formation
process \citep{Jura85,Gail87,Gail99,Ferrarotti06}, but meteoritic evidence
shows that grains condense around nucleation seeds based on titanium
\citep{Bernatowicz91}, which in oxygenous environments are coated by
aluminium-oxides and then by silicates \citep{Vollmer06}. Observations in
47\,Tucanae show indeed that the brightest, AGB star has formed silicate dust
\citep{Vanloon06c}, but that fainter stars generally show features from
aluminium-oxygen bonds \citep{Lebzelter06}.

Although carbon AGB stars produce their own main dust condensate, carbon, and
\citet{Vanloon99} \citep[and][]{Matsuura05,Sloan06,Zijlstra06,Lagadec07a} show
that metal-poor carbon stars have very strong molecular bands because of a
higher availability of carbon, there is no evidence suggesting that metal-poor
carbon stars have a higher dust content. The molecular mass-loss rates are
consistent with the dust mass-loss rates \citep{Matsuura06} for $\psi\propto
Z_{\rm init}$, which was found initially from analysis of dust optical depths
\citep{Vanloon00}. This is further supported by a comparison of the available
amount of silicon with the strength of the silicon-carbide features in carbon
stars in the Fornax dwarf galaxy \citep{Matsuura07}, and a low dust content
seen in magellanic carbon PNe \citep{Stanghellini07}.

Titanium (or zirconium or silicon) is pivotal as a seed for carbonaceous dust
to nucleate onto, so the number density of grains is $n_{\rm grains}\propto
Z_{\rm init}$. Grain growth is determined by $n_{\rm seeds}\times n_{\rm
condensates}$. If more carbon is available, larger molecules form such as
acetylene \citep{Vanloon06b,Matsuura06} or even macromolecules. The {\it
number density} of carbonaceous molecules may thus not be that dissimilar
between carbon stars of different $Z_{\rm init}$, and the dust:gas ratio may
depend mostly on $n_{\rm seeds}$ and thus $\psi\propto Z_{\rm init}$.

\subsection{The r\^ole of pulsation}

\citet{Paczynski68} established a link between Miras and PN ejection, and
thick dust shells are always seen in conjunction with strong pulsation in the
fundamental mode \citep{Jura86}. Pulsation is believed to be the initial stage
in launching a wind, with dust formation providing the final stage in driving
it away \citep{Bowen91,Wachter02}.

%
\begin{figure}[t!]
\resizebox{\hsize}{!}{\includegraphics[clip=true]{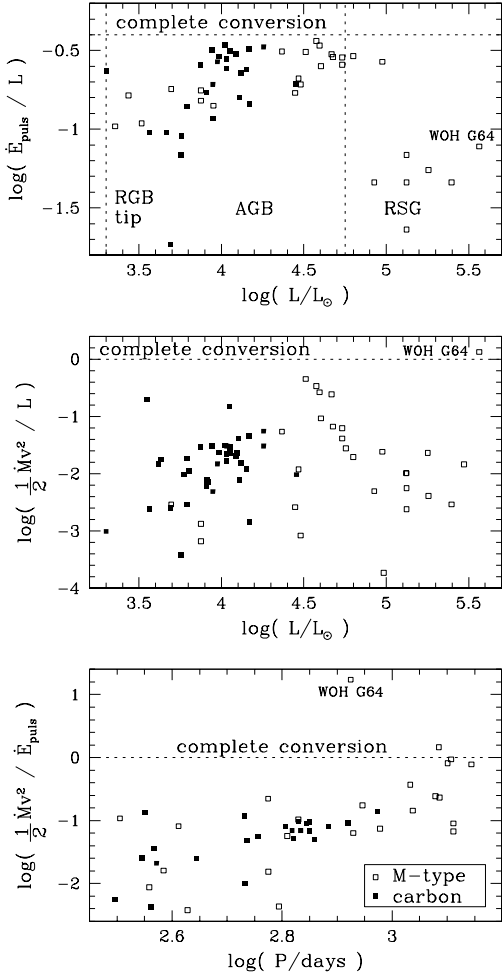}}
\caption{\footnotesize The mean energy rate involved in pulsational expansion
as a fraction of the luminosity, versus luminosity (top), and the kinetic
energy in the ejecta compared to luminosity, versus luminosity (centre)
and compared to the pulsational energy rate, versus pulsation period (bottom).
See text for details.
}
\label{fig1}
\end{figure}

To quantify the potential of pulsation to drive a wind, \citet{Vanloon02}
introduced the mean energy rate involved in pulsation, taking the K-band
amplitude as a proxy for a sinusoidal bolometric light modulation
\citep[cf.][]{Vanloon06b}. Using data in the LMC
\citep{Whitelock03,Vanloon99b,Vanloon05}, the pulsation of AGB stars appears
to saturate just below the maximum attainable conversion of photons into
mantle expansion (Fig. \ref{fig1}, top), whilst RSGs clearly pulsate less
strongly. Nonetheless, the efficiency with which mechanical energy is
transferred from pulsation to a wind is a smooth function of pulsation period
and, as before, indistinguishable for M-type and carbon AGB stars (Fig.
\ref{fig1}, bottom), remaining significantly less than 100\% for all but the
most extreme OH/IR stars. In general, the radiation field is found to contain
10-1000 times more energy than required to support the wind (Fig. \ref{fig1},
centre), independent of stellar luminosity, suggesting the mass loss is
determined more critically by the pulsation than by the radiation.

There exists a transition regime between stars where radial pulsation is
unimportant, and Miras. These semi-regular variables were found to have low
mass-loss rates as derived from IR emission, $\dot{M}\sim10^{-7}$ M$_\odot$
yr$^{-1}$ or less \citep{Knapp98}. On the other hand, CO measurements of
semi-regulars compared to Miras suggested that mass-loss rates are unaffected
by the pulsation mode \citep{Kahane94} --- perhaps semi-regulars are simply
less dusty. At pulsation periods $P>200$ days the mass-loss rates are higher
and increase with increasing period \citep{Knapp98}. This is partially due to
more luminous (bigger) stars having longer periods as well as more radiative
momentum to drive the wind. Similarly, the pulsation period is longer for
cooler (bigger) stars which form dust more easily as it can form closer to the
star where the density is higher. But some stars with ``short'' periods exist
which are cool (such as EP\,Aquarii, R\,Leonis, or W\,Hydrae) but have little
dust. The short pulsation cycle times and smaller amplitudes of semi-regular
variables pulsating in an overtone may leave less time for dust to form, in a
weaker pulsation shock.

Some semi-regulars (e.g., EP\,Aqr) show both a fast and a slow component in
their wind \citep{Knapp98}. This could indicate some instability in the wind
driving mechanism inherent to the pulsational transition regime.

\subsection{The r\^ole of chromospheres}

The chromospherically active, early M-type RSG Betelgeuse is a famous example
of a star which is relatively warm, has little dust and does not pulsate very
strongly, but which nevertheless has $\dot{M}\sim10^{-5}$ M$_\odot$ yr$^{-1}$
\citep{Vanloon05}. Could mass loss from AGB stars be driven by a chromosphere
too?

\citet{Judge91} show that chromospheres and dust-driven winds carry a similar
energy flux as fraction of the bolometric luminosity, $\sim10^{-6}-10^{-4}$,
with little evidence for anything other than a seamless transition. Hence,
they suggested that it does not really matter what the mechanism for driving
the wind is, as plenty of energy is available for a host of mechanisms to
operate, one of which may (but need not) dominate. They suggest that the
ability to lose mass at a given rate is set principally by the depth of the
gravitational potential well, i.e.\ the surface gravity.

\citet{Mcdonald07} analysed the optical line profiles of stars above and below
the RGB-tip in globular clusters; some, but not all, of these stars
have dust. Pulsation shocks were most clearly visible in the H$\alpha$ line
profiles of the strong pulsators, which tend to be the most luminous stars and
likely on the AGB. The mass-loss rates estimated from the absorption line
profiles agreed with those derived from the IR emission \citep{Origlia02}, as
well as with the heuristic model for a circumstellar origin of the H$\alpha$
emission wings \citep{Cohen76}, but they were an order of magnitude higher
than Reimers' Law predicts.

Schr\"oder \& Cuntz (2005) modified Reimers' Law with a heuristic
argument for the calibration in terms of stellar temperature. This would cause
metal-poor (warmer) stars to lose mass at a higher rate. This could cause
bluer horizontal branches. On the other hand, van Loon et al.\ (2007) suggest
metal-poor stars in $\omega$\,Centauri lose slightly less mass on the RGB and
thus become more often post-AGB stars than their metal-richer siblings. This
would also explain the rare presence of a PN in the very metal-poor globular
cluster, M\,15.

\section{Mass loss and AGB evolution}

Mass loss truncates the growth of the core and hence AGB evolution and the
number of thermal pulses. \citet{Vassiliadis93} show that mass loss according
to their formula kicks in more suddenly and later in the AGB evolution than
Reimers' Law, as it depends more extremely on $L$ and $T_{\rm eff}$ --- or
rather radius, as they parameterised it as $\dot{M}(P)$. Mass loss also
reduces the mantle mass and density and thus the effect of $3^{\rm rd}$
dredge-up. This could cause Hot Bottom Burning to stop, allowing for a final
thermal pulse to convert a massive oxygen-rich AGB star into a carbon star
\citep{Vanloon98,Frost98}. On the other hand, I have argued previously that
mass loss will need to become very much less efficient to affect the IFM
relation.

Super-AGB stars, which ignite carbon, are not well represented or recognised
in current samples, which might be understood if they do not become as cool
and hence dust-enshrouded as slightly less massive shell-burning AGB stars or
slightly more massive core-helium burning supergiants. If this means they do
not lose mass very fast then this could prolong their life and thus facilitate
an electron-capture supernova end, rather than leaving an oxygen-neon white
dwarf \citep[cf.][]{Siess07}.

Mass loss causes an AGB star to expand, which may facilitate mass loss. The
pulsation period could then reach the thermal timescale, in which case the
star will adjust itself continuously to the new configuration
\citep{Vanloon02}. It is unclear what effects this will cause.
 
\section{Critical measurements to make}

Measurements which can (soon) be made, and which would greatly advance
understanding of AGB mass-loss, include the following:
\begin{itemize}
\item[1]{Measure the IFM relation at $<0.1$ Z$_\odot$;}
\item[2]{Measure the wind speed and dust:gas ratio in metal-poor carbon
stars;}
\item[3]{Measure chromospherically-driven winds;}
\item[4]{Correlate mass-loss rates with gravities;}
\item[5]{Spectro-interferometric monitoring of the pulsating atmosphere and
dust formation region, (near-)contemporaneous with interferometric maser
observations;}
\item[6]{Reconstruct the mass-loss history through IR or (sub)mm observations
of envelopes;}
\item[7]{Population synthesis of Local Group AGB star populations to {\it
derive} a prescription for mass-loss rate, dust production and yields as
function of stellar parameters and time.}
\end{itemize}

\begin{acknowledgements}
To the organisers for inviting me to present this review, the editors (and
Joana) for their infinite patience, and everybody who was there for a very
pleasant and stimulating atmosphere: Thank You!
\end{acknowledgements}

\bibliographystyle{aa}

\end{document}